\documentclass[12pt]{article}

\usepackage{graphicx}

\begin{document}
\begin{center}

{\bf Notes on Born$-$Infeld-like modified gravity }\\
\vspace{5mm}
 S. I. Kruglov
 \footnote{serguei.krouglov@utoronto.ca}

\vspace{5mm}
\textit{Department of Chemical and Physical Sciences, University of Toronto,\\
3359 Mississauga Rd. North, Mississauga, Ontario, Canada L5L 1C6}
\end{center}

\begin{abstract}
We investigate the modified $F(R)$ gravity theory with the function
$F(R) = (1-\sqrt{1-2\lambda R-\sigma (\lambda R)^2})/\lambda$.
The action is converted into Einstein$-$Hilbert action at small values of $\lambda$ and $\sigma$. The local tests give a bound on the parameters, $\lambda(1+\sigma)\leq 2\times 10^{-6}$ cm$^2$. The Jordan and Einstein frames are considered, the potential, and the mass of the scalar field were obtained. The constant curvature solutions of the model are found. It was demonstrated that the de Sitter space is unstable but a solution with zero Ricci scalar is stable. The cosmological parameters of the model are evaluated. Critical points of autonomous equations are obtained and described.
\end{abstract}


\section{Introduction}

One of ways to explain the inflation and the present time of the Universe acceleration is to modify the Einstein$-$Hilbert (EH) action of general relativity (GR) theory. Here we consider the $F(R)$ gravity model replacing the Ricci scalar $R$ in EH action by the function $F(R)$. Such $F(R)$ gravity model can be an alternative to $\Lambda$-Cold Dark Matter ($\Lambda$CDM) model where the cosmic acceleration appears due to modified gravity. Thus, instead of the introduction of the cosmological constant $\Lambda$ (having the problem with the explanation of the smallness of $\Lambda$) to describe dark energy (DE), we consider new gravitational physics.
The requirement of classical and quantum stability leads to the conditions $F'(R)>0$, $F''(R)>0$ (Appleby et al. 2010), where the primes mean the derivatives with respect to the argument. These conditions do not fix the function and, therefore, there are various suggestions in the form of the function $F(R)$ in the literature.
It should be mentioned that the first successful models of $F(R)$ gravity were given in Hu (2007), Appleby and Battye (2007), and Starobinsky (2007).
The modified gravitational theories $f(R,T)$ with non-minimal curvature matter coupling, where the gravitational Lagrangian is given by an arbitrary function of the Ricci scalar $R$ and of the trace of the stress-energy tensor $T$, were considered by Harko (2011), Sharif (2014), Zubair (2015), Noureen (2015). It was shown the possibility of the transition from decelerating to accelerating phase in some f(R,T) models.

In this paper we investigate the Born$-$Infeld (BI) type Lagrangian with the particular function $F(R)= \left(1-\sqrt{1-2\lambda R-\sigma\left(\lambda R\right)^2}\right)/ \lambda$ introducing two scales. In BI electrodynamics there are no divergences connected with point-like charges and the self-energy is finite (Born and Infeld (1934a), Born and Infeld (1934b), Born and Infeld (1935), Plebanski (1970)). In addition, BI type action appears naturally within the string theory. Thus, the low energy D-brane dynamics is governed
by a BI type action (Fradkin and Tseytlin (1985)). These two attractive aspects of BI type theories are the motivation to consider BI-like gravity. In Kruglov (2010) we have considered modified BI electrodynamics with two constants. The model under consideration is the gravitational analog of generalized BI electrodynamics with two scales.
It should be also mentioned that there are difficulties to quantize $F(R)$ gravity because it is the higher derivative (HD) theory. In HD theories there are additional degrees of freedom and ghosts are present so that unitarity of the
theory is questionable. In addition, corrections due to one-loop divergences, introduced by renormalization, contain a scalar curvature squared ($R^2$) and the Ricci tensor squared ($R_{\mu\nu}R^{\mu\nu}$).
As a result, $F(R)$ gravity theories are not renormalizable. At the same time, $F(R)$ gravity is the phenomenological model, and can give a description of the Universe evolution including the inflation and the late-time acceleration, modifies gravitational physics, and is an alternative to the $\Lambda$CDM model. The first model including $R^2$ term in the Lagrangian, and describing the self-consistent inflation, was given in Starobinsky (2007).

The paper is organized as follows. In Sec. 2, we consider a model of $F(R)$ gravity with the BI-like Lagrangian density with two scales. A bound on the parameters $\lambda$ (with the dimension
(length)$^2$) and $\sigma$ (the dimensionless parameter) is obtained. Constant curvature solutions corresponding the de Sitter space are obtained. In Sec. 3, the scalar-tensor form of the model is studied, the potential of the scalar degree of freedom and the mass are found, and the plots of the functions $\phi(\lambda R)$, $V(\lambda R)$, and $m^2_{\phi}(\lambda R)$ are given for $\sigma=-0.9$. We show that the de Sitter phase is unstable and the flat space (a solution with the zero curvature scalar) is stable. The slow-roll cosmological parameters of the model under consideration are evaluated and the plots of functions $\epsilon(\lambda R)$, $\eta(\lambda R)$ are given in Sec. 4. In Sec. 5 critical points of autonomous equations are investigated. The function m(r) characterizing the deviation from the $\Lambda$CDM model is evaluated and the plot is presented. A particular case of the model with the parameter $\sigma=0$ is studied in subsection 5.1 in details. The results obtained are discussed in Sec. 6.

The Minkowski metric $\eta_{\mu\nu}$=diag(-1, 1, 1, 1) is used and we assume $c$=$\hbar$=1.

\section{The Model}

We propose the modified $F(R)$ gravity model with the Lagrangian density
\begin{equation}
{\cal L}=\frac{1}{2\kappa^2}F(R)=\frac{1}{2\kappa^2}\frac{1}{\lambda}\left(1-\sqrt{1-2\lambda R-\sigma\left(\lambda R\right)^2}\right),
\label{1}
\end{equation}
where $\kappa=\sqrt{8\pi}m_{Pl}^{-1}$, $m_{Pl}$ is the Planck mass, $\lambda$ has the dimension of (length)$^2$ and $\sigma$ is dimensionless parameter. The model with the Lagrangian density (1) is the generalized form of the model discussed in (Kruglov (2013)). This extension, containing the additional parameter $\sigma$, allows us to investigate how the Universe evolution and cosmological parameters depend on $\sigma$. This can help to find some values of $\sigma$ for a realistic and viable model.
The action is given by
\begin{equation}
S=\int d^4x\sqrt{-g}\left({\cal L}+{\cal L}_m\right)=\int d^4x\sqrt{-g}\left[\frac{1}{2\kappa^2}F(R)+{\cal L}_m\right],
\label{2}
\end{equation}
were ${\cal L}_m$ is the matter Lagrangian density.
In the works of Comelli and Dolgov (2004), Comelli (2005), Quiros et al. (2009), Garcia-Salcedo (2010) the BI type model involving the Gauss-Bonnet term $\mathcal{G}=R^2-4R_{\mu\nu}R^{\mu\nu}+R_{\mu\nu\alpha\beta}R^{\mu\nu\alpha\beta}$, instead of $R^2$-term in (1), was considered. But that model belongs to the $F(R,R_{\mu\nu}R^{\mu\nu},R_{\mu\nu\alpha\beta}R^{\mu\nu\alpha\beta})$ type model. In EH action only the scalar curvature $R$ is present, and therefore, we neglect the higher order invariants $R_{\mu\nu}R^{\mu\nu}$, $R_{\mu\nu\alpha\beta}R^{\mu\nu\alpha\beta}$ and consider $F(R)$ gravity model.
It follows from Eq. (1) that at $\sigma=-1$, we have $F(R)=R$ and one comes to the EH action. Therefore, we imply that the unitless parameter $\sigma$ is in the order of $1$ and is close to ($-1$) to recover GR at low curvature regime.
At a particular case $\sigma=0$ different aspects of the model were investigated in Comelli (2005), Garcia-Salcedo (2010), Kruglov (2013). Other variants of BI-type gravity were considered in Deser and Gibbons (1998), Gates, Jr. and Ketov (2001), Wohlfarth (2004a), Wohlfarth (2004b), Nieto (2004), Gullu et al. (2010),  Ba\~{n}ados and Ferreira (2010), Quiros and Urena-Lopez (2010), Pani et al. (2011), Herdeiro et al. (2011), Fabris et al. (2012), Makarenko et al. (2014).
Since the function in Eq. (1) should be real, one has the restriction $2\lambda R +\sigma(\lambda R)^2\leq 1$. The Taylor series for small values of $\lambda R$ gives
\begin{equation}
F(R)= R+\frac{1}{2}\lambda\left(1+\sigma\right) R^2+....
\label{3}
\end{equation}
Thus, at small values of the constants $\lambda$ and $\sigma$ introduced, one comes to Starobinsky's model (Starobinsky 2007), that leads to the self-consistent inflation (Appleby et al. 2010).
The model under consideration satisfies observational data at a bound on $\lambda$ and $\sigma$ because GR passes local tests. From laboratory experiment (Kapner et al. 2007), (N\"{a}f and Jetzer 2010), (Berry and Gair 2011), (Eingorn and Zhuk 2011) the restriction on the function is $F''(0)\leq 2\times 10^{-6}$ cm$^2$. Then, we obtain from Eq. (3) the bound on the parameters $\lambda$, $\sigma$:
\begin{equation}
\lambda\left(1+\sigma\right)\leq 2\times 10^{-6} \mbox{cm}^2.
\label{4}
\end{equation}
Since the Taylor series (3) contains all powers in Ricci curvature $R$ at $\lambda R < 1$, the model under consideration can give nontrivial description of the Universe evolution.

\subsection{Constant Curvature Solutions}

In the case when the Ricci scalar is a constant $R=R_0$ equations of motion give (Barrow and Ottewill 1983)
\begin{equation}
2F(R_0)-R_0F'(R_0)=0.
\label{5}
\end{equation}
From Eq. (1), one finds
\begin{equation}
\frac{2}{\lambda}\left(1-\sqrt{1-2\lambda R_0-\sigma(\lambda R_0)^2}\right)=\frac{R_0\left(1+\sigma \lambda R_0\right)}{\sqrt{1-2\lambda R_0-\sigma(\lambda R_0)^2}}.
\label{6}
\end{equation}
Eq.(6) can be written as
\begin{equation}
x\left(\sigma^2x^3+6\sigma x^2+9x-4 \right)=0,
\label{7}
\end{equation}
where $x=\lambda R_0$. The trivial solution to Eq. (7) $x=0$ corresponds to flat spacetime. For $1>\sigma >-1$ cubic Eq.(7) possesses nontrivial real solutions.
From Eq. (7) one can find the value of $\sigma$ as a function of $x$ (we use only one root obeying $1>\sigma >-1$)
\begin{equation}
\sigma=\frac{2-3\sqrt{x}}{x^{3/2}}.
\label{8}
\end{equation}
Below in Table 1 there are approximate values of $x$, which are the roots of Eq. (7), for different values of the parameter $\sigma$.
The plot of the function $\sigma$ vs $\lambda R$, Eq. (8), is represented by Fig.1.
\begin{table}[ht]
\caption{Approximate solutions to Eq. (7)}
\centering
\begin{tabular}{c c c c c c c c c c c c c c}
\hline \hline 
$\sigma$ & -0.1 & -0.3 & -0.5 & -0.7& -0.9 & 0 & 0.1 & 0.2 & 0.3 & 0.4& 0.5 \\[0.5ex] 
\hline 
x & 0.46 & 0.49 & 0.54 & 0.60 & 0.73 & 4/9 & 0.43 & 0.42 & 0.41 & 0.40& 0.39 \\
[1ex] 
\hline 
\end{tabular}
\label{table:Approx}
\end{table}
\begin{figure}[h]
\includegraphics[height=4.0in,width=4.0in]{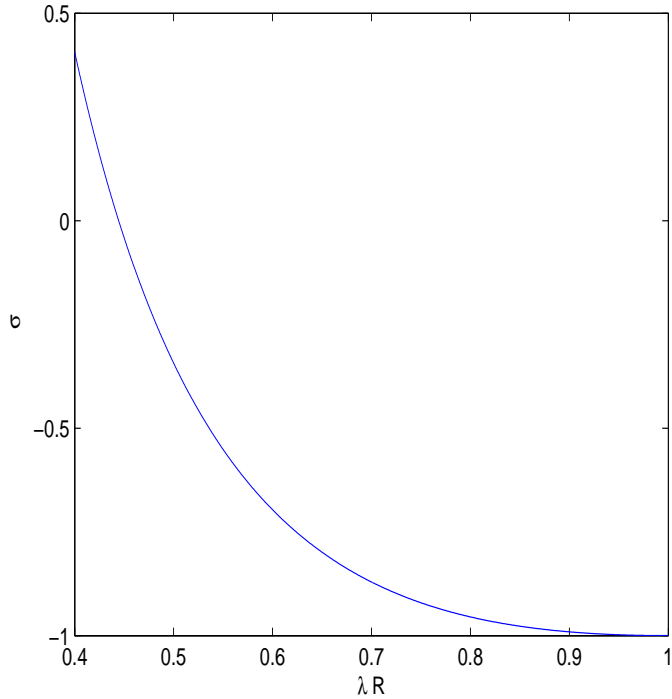}
\caption{\label{fig.1}The function $\sigma$ vs $\lambda R$ corresponding the constant curvature solutions. }
\end{figure}
The conditions of classical stability $F'(R)>0$ and quantum stability $F''(R)> 0$ (Appleby et al. 2010) lead to the restrictions
\begin{equation}
F'(R)=\frac{\left(1+\sigma \lambda R\right)}{\sqrt{1-2\lambda R-\sigma(\lambda R)^2}}>0,
\label{9}
\end{equation}
\begin{equation}
F''(R)=\frac{\lambda\left(1+\sigma \right)}{\left[1-2\lambda R-\sigma(\lambda R)^2\right]^{3/2}}>0.
\label{10}
\end{equation}
Inequalities in Eqs. (9),(10) are satisfied at $\lambda>0$, $\sigma>-1$ ($2\lambda R+\sigma(\lambda R)^2\leq 1$ and the $F(R)$ is a real function). One can verify that conditions (9),(10) are satisfied for constant curvature solutions obtained (see Table 1). As a result, nontrivial solutions to Eq. (7) lead to the Schwarzschild-de Sitter spacetime and the trivial solution $R=0$ corresponds to the Minkowski spacetime. If the condition $F'(R_0)/F''(R_0)>R_0$ holds (M\"{u}ller 1988), it can describe DE which is future stable. This leads to
\begin{equation}
x\left(\sigma^2x^2+3\sigma x+3\right)<1.
\label{11}
\end{equation}
The trivial solution $R_0=0$ obeys the requirement (11) and it is stable. One may verify that nontrivial solutions to Eq. (7) represented in Table 1 do not satisfy Eq. (11). Therefore, constant curvature solutions (in Table 1) give unstable de Sitter spacetime and can describe the inflation. Below we show that constant curvature solutions correspond to the maximum of the effective potential in Einstein's frame. Thus, the model suggested mimics DE for the spacetime without matter.

\section{The Scalar-Tensor Formulation of the Theory}

\begin{figure}[h]
\includegraphics[height=4.0in,width=4.0in]{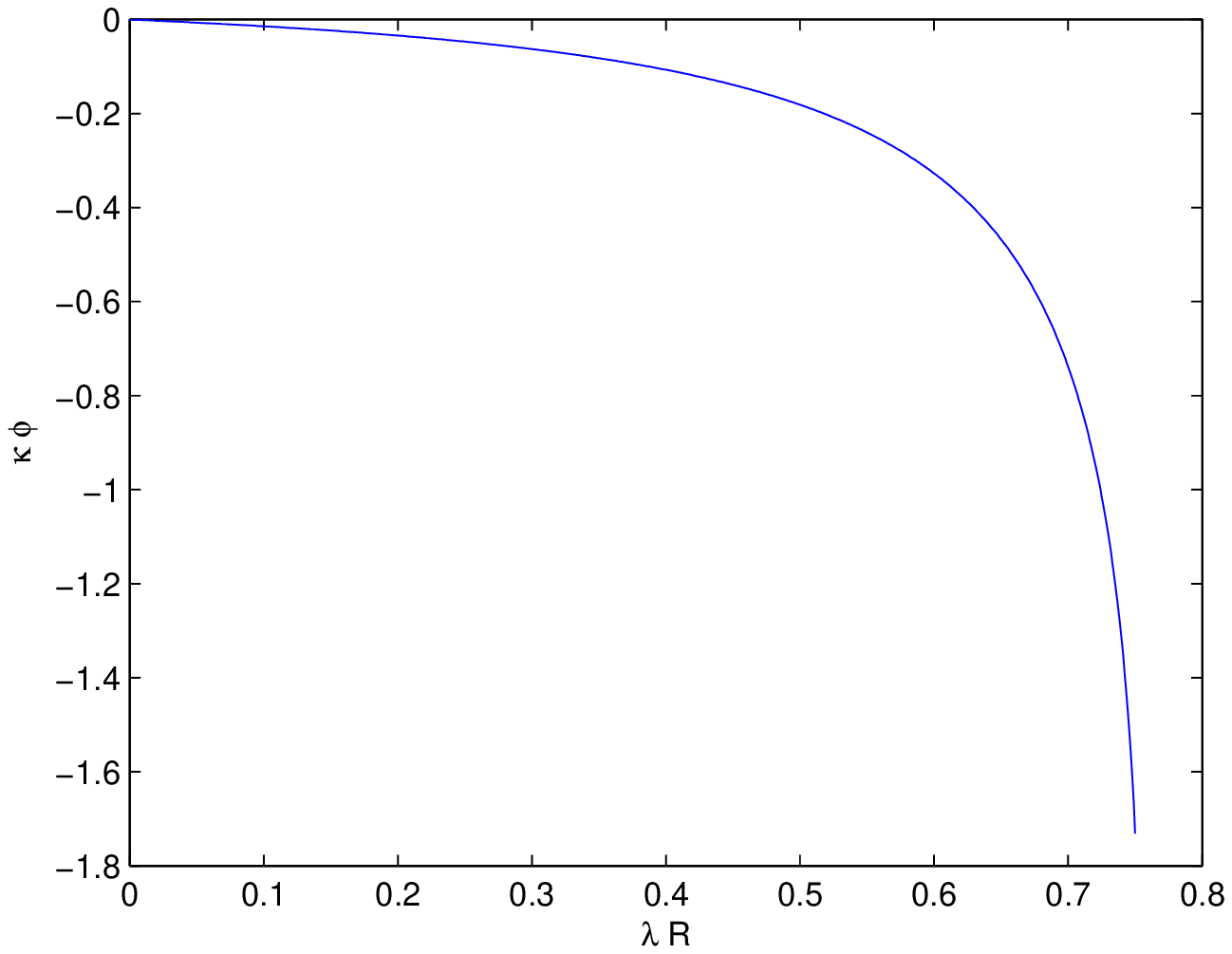}
\caption{\label{fig.2}The function  $\kappa\phi$ ($\sigma=-0.9$) vs $\lambda R$. }
\end{figure}

\begin{figure}[h]
\includegraphics[height=4.0in,width=4.0in]{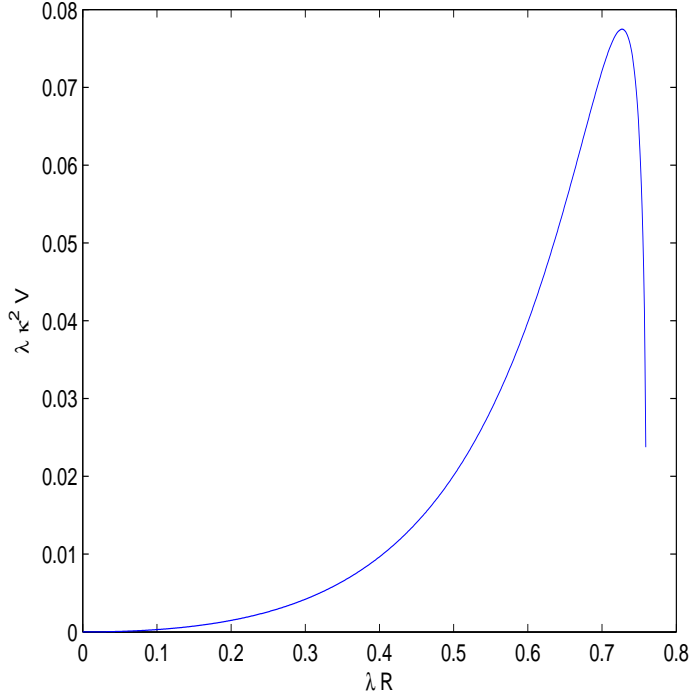}
\caption{\label{fig.3}The function $\lambda\kappa^2V$ ($\sigma=-0.9$) vs $\lambda R$. There is the maximum at $\lambda R\approx 0.73$ and the minimum at $R=0$.}
\end{figure}

The modified $F(R)$ gravity model in the Jordan frame can be represented in the scalar-tensor form corresponding the Einstein frame. Thus, we perform the conformal transformation of the metric (Magnano and Sokolowski 1994)
\begin{equation}
\widetilde{g}_{\mu\nu} =F'(R)g_{\mu\nu}=\frac{\left(1+\sigma \lambda R\right)g_{\mu\nu}}{\sqrt{1-2\lambda R-\sigma(\lambda R)^2}},
\label{12}
\end{equation}
and Eq. (1) takes the form of the Lagrangian density corresponding the scalar-tensor theory of gravity
\begin{equation}
{\cal L}=\frac{1}{2\kappa^2}\widetilde{R}-\frac{1}{2}\widetilde{g}^{\mu\nu}
\nabla_\mu\phi\nabla_\nu\phi-V(\phi).
\label{13}
\end{equation}
The scalar curvature $\widetilde{R}$ in Einstein's frame is calculated by the new metric (12). The scalar field $\phi$ and the potential $V(\phi)$ are given by
\begin{equation}
\phi(R)=-\frac{\sqrt{3}}{\sqrt{2}\kappa}\ln F'(R)=\frac{\sqrt{3}}{\sqrt{2}\kappa}\ln \frac{\sqrt{1-2\lambda R-\sigma (\lambda R)^2}}{1+\sigma\lambda R},
\label{14}
\end{equation}
\[
V(R)=\frac{RF'(R)-F(R)}{2\kappa^2F'^2(R)}
\]
\vspace{-7mm}
\begin{equation}
\label{15}
\end{equation}
\vspace{-7mm}
\[
=\frac{\left(1-\lambda R\right)\sqrt{1-2\lambda R-\sigma (\lambda R)^2}+2\lambda R+\sigma(\lambda R)^2-1}{2\lambda\kappa^2\left(1+\sigma\lambda R\right)^2}.
\]
The plots of the functions $\phi(R)$ and $V(R)$ at $\sigma=-0.9$ are represented in Fig. \ref{fig.2} and Fig. \ref{fig.3}, respectively.
From Eq. (15) we find the extremum of the potential
\begin{equation}
\frac{dV}{dR}=\frac{F''(R)\left[2F(R)-RF'(R)\right]}{2\kappa^2F^{'3}}=0.
\label{16}
\end{equation}
From Eqs. (5),(16) we make a conclusion that the constant curvature solutions to Eq. (7) correspond to the extremum of the potential. The potential function (15) possesses the minimum at $\phi=0$ and the maximum is given by Eq. (7) and by Fig. 1 for different parameters $\sigma$. The Minkowski spacetime ($R=0$) is the stable state and the states with the curvatures obeying Eq. (7) are unstable. From potential (15), one finds the mass squared of a scalar state
\[
m_\phi^2=\frac{d^2V}{d\phi^2} =\frac{1}{3}\left(\frac{1}{F''(R)}+\frac{R}{F'(R)}-\frac{4F(R)}{F^{'2}(R)}\right)
\]
\begin{equation}
=\frac{1}{3\lambda}\biggl\{\frac{4x^2-3x+5+\sigma^2x(x^3+x+4)+\sigma(4x^3-x^2+5x+4)}
{(1+\sigma)\sqrt{1-2x-\sigma x^2}}
\label{17}
\end{equation}
\[
-\frac{4(1+\sigma x)}{(1-2x-\sigma x^2)}\biggr\}.
\]
The plot of the function $\lambda m_\phi^2$ vs $x=\lambda R$ is given by Fig. 4.
\begin{figure}[h]
\includegraphics[height=4.0in,width=4.0in]{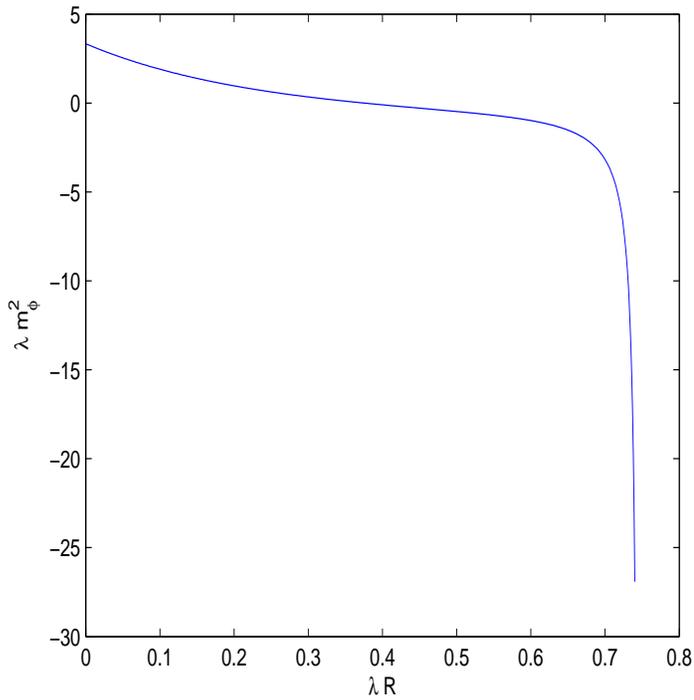}
\caption{\label{fig.4}The function $\lambda m^2_\phi$ ($\sigma=-0.9$) vs $\lambda R$. }
\end{figure}
It is seen from Fig. 4 that for the constant curvature solutions given by Table 1 the values of $m^2_\phi$ are negative ($m^2_\phi<0$) that again indicates on instability of states corresponding to solutions of Eq. (7).
The criterion of the stability of the de Sitter solution in F(R) gravity models was first obtained in M\"{u}ller et al. (1988).  Since the constant $\lambda$ is small the squared mass $m_\varphi^2$ is big and corrections to the Newton law are negligible.

\section{The Slow-Roll Cosmological Parameters of the Model}

The requirement that corrections of $F(R)$ gravity model are small compared to GR for $R\gg R_0$, where $R_0$ is a curvature at the present time, gives (Appleby et al. (2010))
\begin{equation}
\mid F(R)-R\mid \ll R,~\mid F'(R)-1\mid \ll 1,~\mid RF''(R)\mid\ll 1.
\label{18}
\end{equation}
With the help of Eq. (3), in the liner approximation, we obtain from Eq. (18) the restriction $\lambda R\ll 1/(1+\sigma)$. Thus, for $\sigma=-0.9$, we obtain $\lambda R\ll 10$ which is satisfied (see figures).\\
Let us consider the slow-roll parameters which are given by (Liddle and Lyth 2000)
\begin{equation}
\epsilon(\phi)=\frac{1}{2}M_{Pl}^2\left(\frac{V'(\phi)}{V(\phi)}\right)^2,~~~~\eta(\phi)=M_{Pl}^2\frac{V''(\phi)}{V(\phi)},
\label{19}
\end{equation}
were the reduced Planck mass is $M_{Pl}=\kappa^{-1}$. From Eqs. (15),(17) we obtain the slow-roll parameters expressed via the curvature
\begin{equation}
\epsilon=\frac{1}{3}\left[\frac{RF'(R)-2F(R)}{RF'(R)-F(R)}\right]^2=\frac{1}{3}\left[\frac{2-3x-\sigma x^2-2\sqrt{1-2x-\sigma x^2}}{1-x-\sqrt{1-2x-\sigma x^2}}\right]^2,
\label{20}
\end{equation}
\[
\eta=\frac{2}{3}\left[\frac{F^{'2}(R)+F''(R)\left[RF'(R)-4F(R)\right]}{F''(R)\left[RF'(R)-F(R)\right]}\right]
\]
\begin{equation}
=\frac{2\left[5-9x+4\sigma-\sigma x\left(5+2\sigma x+8x+4\sigma x^2+\sigma^2 x^3\right)\right]}{3\left(1+\sigma\right)\left(1-x-\sqrt{1-2x-\sigma x^2}\right)}
\label{21}
\end{equation}
\[
-\frac{8\sqrt{1-2x-\sigma x^2}}{3\left(1-x-\sqrt{1-2x-\sigma x^2}\right)},
\]
where $x=\lambda R$. The function $\phi(R)$ is given by Eq. (14) and represented by Fig. 2. The slow-roll approximation is valid when the conditions $\epsilon\ll 1$, $\mid\eta\mid\ll 1$ are satisfied. The plots of the functions $\epsilon$, $\eta$ at $\sigma=-0.9$ are given in Fig. 5 and Fig. 6, respectively.
\begin{figure}[h]
\includegraphics[height=4.0in,width=4.0in]{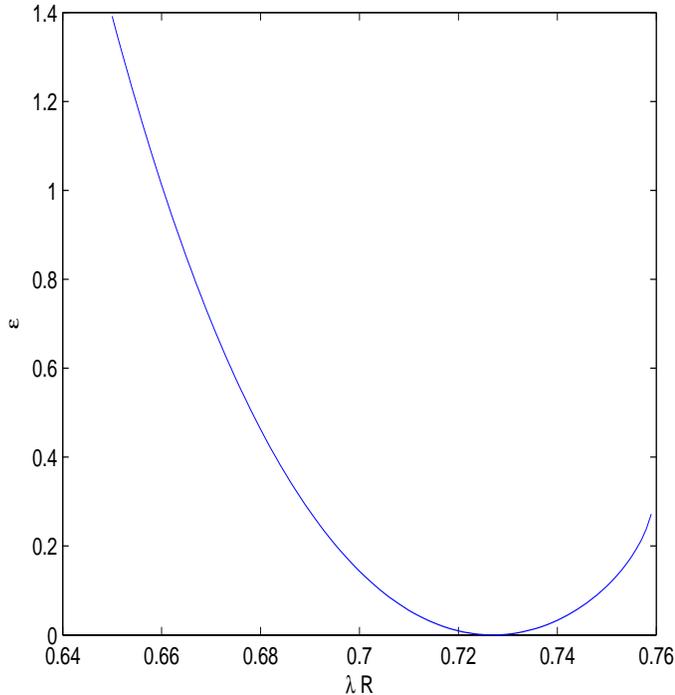}
\caption{\label{fig.5} The function $\epsilon$ vs $\lambda R$ ($\sigma=-0.9$). }
\end{figure}
\begin{figure}[h]
\includegraphics[height=4.0in,width=4.0in]{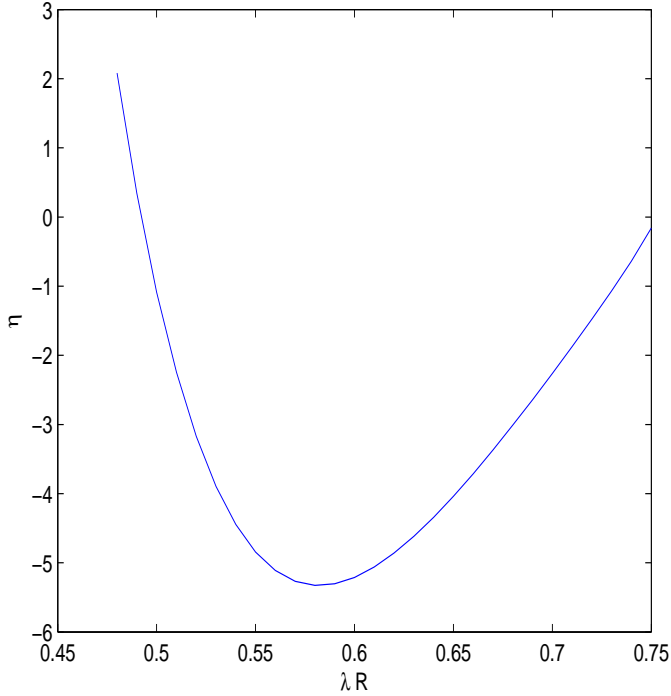}
\caption{\label{fig.6}The function $\eta$ vs $\lambda R$ ($\sigma=-0.9$). }
\end{figure}
The inequality $\epsilon <1$ holds at $\lambda R>0.66$ and $|\eta | <1$ at $\lambda R>0.732$ (and at $0.486<\lambda R<0.499$).
\\
One can calculate the age of the inflation by evaluating the $e$-fold number (Liddle and Lyth (2000))
\begin{equation}
N_e\approx \frac{1}{M_{Pl}^2}\int_{\phi_{end}}^{\phi}\frac{V(\phi)}{V'(\phi)}d\phi.
\label{22}
\end{equation}
The $\phi_{end}$ corresponds to the time at the end of inflation. We find the number of $e$-foldings from Eqs. (14),(15)
\begin{equation}
N_e\approx \frac{3(1+\sigma)}{2}\int_{x_{end}}^x\frac{1-x-y(x)}{y^2(x)(1+\sigma x)\left[2y(x)-2+3x+\sigma x^2\right]}dx,
\label{23}
\end{equation}
were $y(x)=\sqrt{1-2x-\sigma x^2}$, and the value $x_{end}=\beta R_{end}$ corresponds to the time of the end of the inflation when $\epsilon$ or $|\eta |$ are close to $1$. One can verify that even for $x_{end}=0.7$ and $x=0.7269$, we get $N_e\approx 6$ that is not reasonable amount of the inflation (Liddle and Lyth 2000). Thus, the model under consideration describes the inflation but does not reproduce the necessary age of the inflation. Therefore, this model can give an approximated account of the Universe evolution.

\section{Critical Points and Stability}

Let us consider the dimensionless parameters (Amendola et al. 2007)
\begin{equation}
x_1=-\frac{\dot{F}'(R)}{HF'(R)},~x_2=-\frac{F(R)}{6F'(R)H^2},~x_3=\frac{\dot{H}}{H^2}+2,
\label{24}
\end{equation}
\begin{equation}
m=\frac{RF''(R)}{F'(R)},~~~~r=-\frac{RF'(R)}{F(R)}=\frac{x_3}{x_2},
\label{25}
\end{equation}
where $H$ is a Hubble parameter and the dot defines the derivative with respect to the cosmic time and the function $m(r)$ (see Amendola et al. (2007)) characterizes the deviation from the $\Lambda$CDM model. With the help of variables (24) equations of motion in the absence of the radiation ($\rho_{\mbox{rad}}=0$) can be represented as autonomous equations (Amendola et al. (2007)):
\begin{equation}
\frac{dx_i}{dN}=f_i(x_1,x_2,x_3),
\label{26}
\end{equation}
where $i=1,2,3$, $N=\ln a$ is the number of $e$-foldings, and functions $f_i(x_1,x_2,x_3)$ are given by
\[
f_1(x_1,x_2,x_3)=-1-x_3-3x_2+x_1^2-x_1x_3,
\]
\begin{equation}
f_2(x_1,x_2,x_3)=\frac{x_1x_3}{m}-x_2\left(2x_3-4-x_1\right),
\label{27}
\end{equation}
\[
f_3(x_1,x_2,x_3)=-\frac{x_1x_3}{m}-2x_3\left(x_3-2\right).
\]
The critical points for the system of equations can be investigated by the study  of the function $m(r)$. From Eqs. (1),(9) and (10), one finds
\[
m=\frac{x(1+\sigma)}{(1+\sigma x)(1-2x-\sigma x^2)},
\]
\vspace{-7mm}
\begin{equation}
\label{28}
\end{equation}
\vspace{-7mm}
\[
r=-\frac{x(1+\sigma x)}{\sqrt{1-2x-\sigma x^2}-1+2x+\sigma x^2},
\]
where $x=\lambda R$. The plot of the function $m(r)$ is presented in Fig. 7.
\begin{figure}[h]
\includegraphics[height=4.0in,width=4.0in]{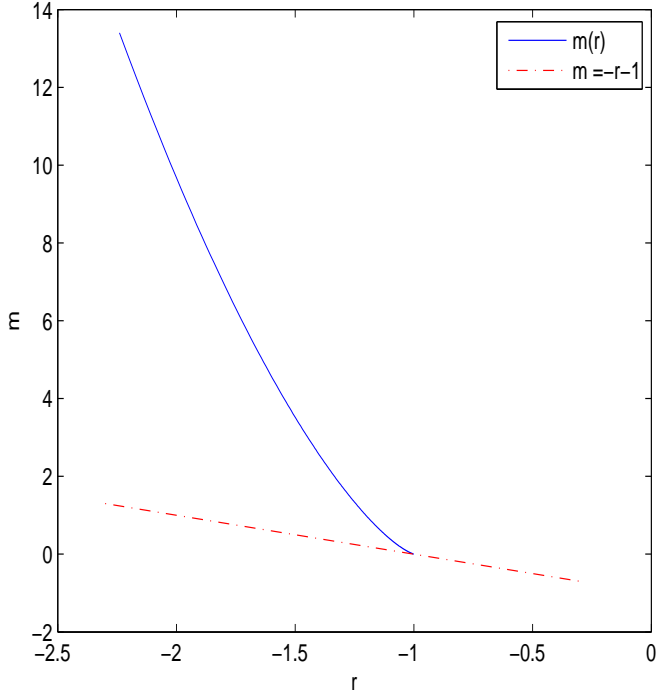}
\caption{\label{fig.7} The function $m(r)$ ($\sigma=-0.9$). }
\end{figure}
The de Sitter point $P_1$ (Amendola et al. (2007)) (in the absence of radiation, $x_4 = 0$) corresponds to the parameters $x_1=0$, $x_2=-1$, $x_3=2$. By virtue of Eqs. (5),(6),(25), one can verify that this point corresponds to the constant curvature solutions ($\dot{H}=0$). The matter energy fraction parameter is given by $\Omega_{\mbox{m}}=1-x_1-x_2-x_3=0$ and the effective equation of state (EoS) parameter $w_{\mbox{eff}}=-1-2\dot{H}/(3H^2)=-1$ which corresponds to DE. In accordance with Fig. 7 we have $1< m(r=-2)$ and, therefore, the constant curvature solution $x \approx 0.73$ ($\sigma=-0.9$) gives unstable the de Sitter spacetime (Amendola et al. (2007)) (it was already mentioned). A viable matter dominated epoch prior to late-time acceleration exists for the critical point $P_5$ (with EoS of a matter era $w_{\mbox{eff}}=0$, $a=a_0t^{2/3}$) with $m\approx 0$, $r\approx -1$ (Amendola et al. (2007)), and for this point $x_3=1/2$. The point $P_5$ belongs to the equation $m=-r-1$ which has only the solution $m=0$, $r=-1$ ($R=0$) in our case. For the existence of the standard matter era
the condition $m'(r=-1)>-1$ should hold (Amendola et al. (2007)). From Eqs. (28) one finds the derivative $m'(r)=(dm/dx)(dx/dr)$:
\begin{equation}
\frac{dm}{dr}=\frac{(1+\sigma)(1+3\sigma x^2+2\sigma^2x^3)\left(y(x)-1\right)^2}{(1+\sigma x)^2y(x)
\left[(1+2\sigma x)y^2(x)\left(y(x)-1\right)+x(1+\sigma x)^2\left(2y(x)-1\right)\right]},
\label{29}
\end{equation}
were $y(x)=\sqrt{1-2x-\sigma x^2}$. From Eq. (29) with the help of L'H\^{o}pital's rule, we obtain the limit: $\lim_{x\rightarrow 0}m'(r)=-2$. Thus, the condition $m'(r=-1)>-1$ does not hold and, therefore, the correct description of the standard matter era in the model is questionable.
As a result, three general conditions, described in (Amendola et al. (2007)) for a successful F(R) model are not satisfied.
Thus, the model under consideration does not lead completely to an acceptable cosmology. But
for a detailed description of the Universe evolution one needs a numerical analysis of autonomous equations (Amendola et al. (2007)).

\subsection{The Particular Case, $\sigma =0$}

\begin{figure}[h]
\includegraphics[height=4.0in,width=4.0in]{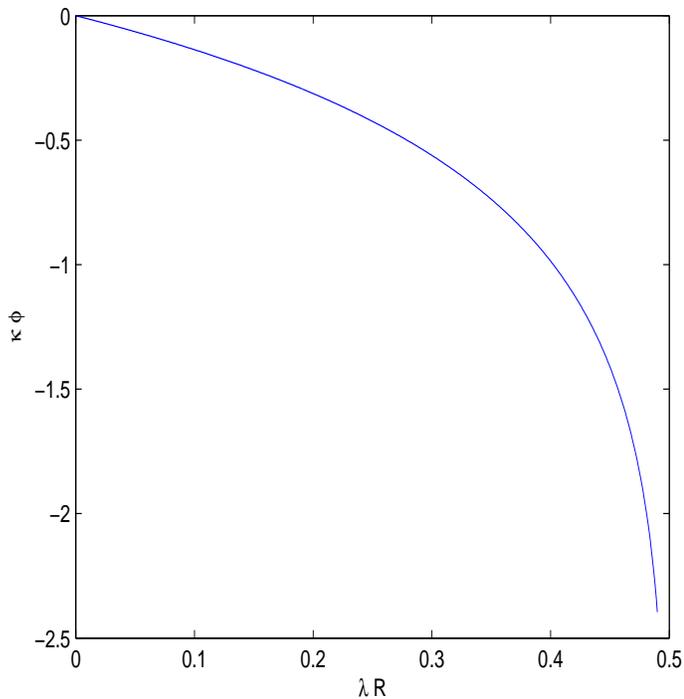}
\caption{\label{fig.8}The function  $\kappa\phi$ ($\sigma=0$) vs $\lambda R$. }
\end{figure}

\begin{figure}[h]
\includegraphics[height=4.0in,width=4.0in]{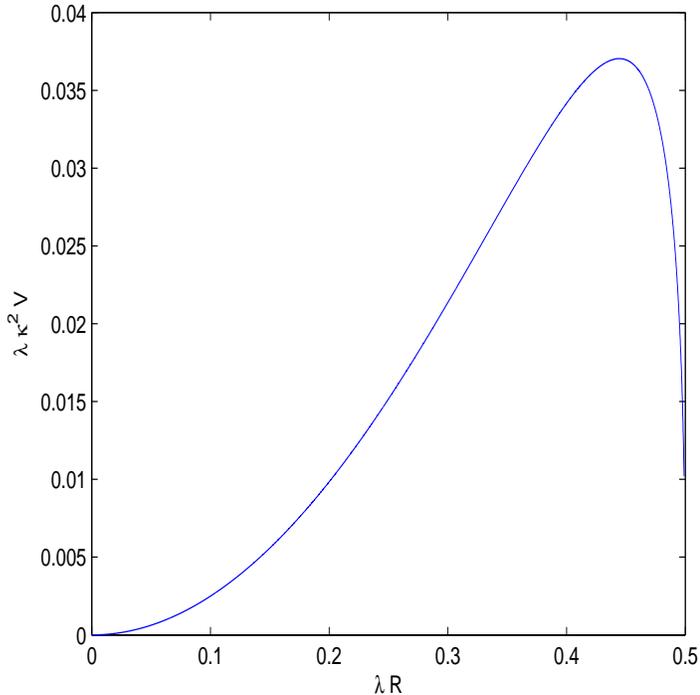}
\caption{\label{fig.9}The function $\lambda\kappa^2V$ ($\sigma=0$) vs $\lambda R$. The maximum is at $\lambda R= 4/9$ and the minimum at $R=0$.}
\end{figure}

\begin{figure}[h]
\includegraphics[height=4.0in,width=4.0in]{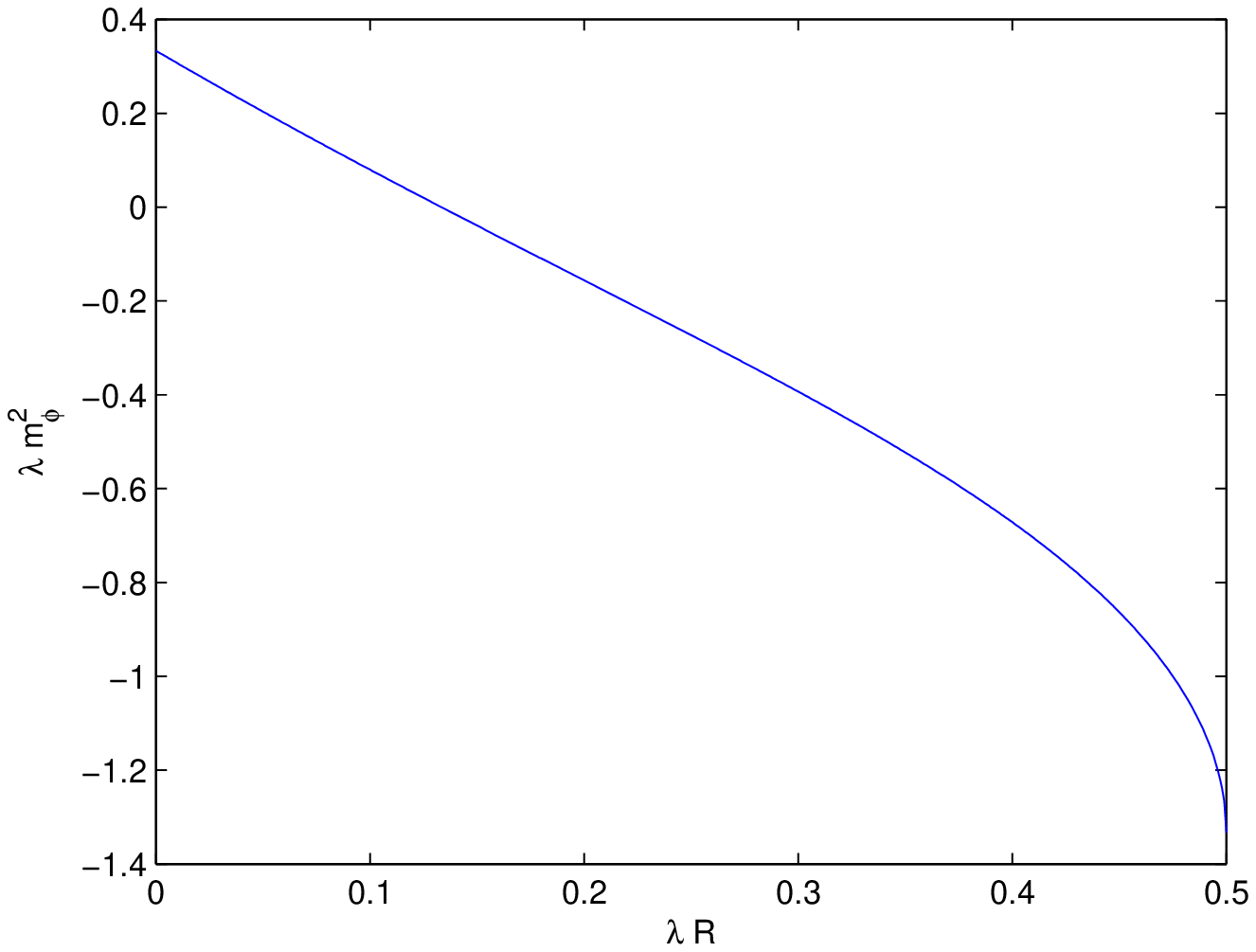}
\caption{\label{fig.10}The function $\lambda m^2_\phi$ ($\sigma=0$) vs $\lambda R$. }
\end{figure}

This case ($\sigma =0$) corresponds to BI procedure of replacing $R$ in EH action by the function $F(R)=(1-\sqrt{1-2\lambda R})/\lambda$ (Comelli 2005), (Quiros et al. 2009), (Garcia-Salcedo et al. 2010), (Kruglov 2013). The plots of the functions $\phi(R)$, $V(R)$, and $m_\phi^2$ given by Eqs. (14),(15) and (17) at $\sigma=0$ are represented in Figs. \ref{fig.8}, \ref{fig.9} and \ref{fig.10}. We also define the EoS parameter $w_{\mbox{eff}}$, the deceleration parameter $q$, and the matter density parameter $\Omega_{\mbox {m}}$ as follows:
\begin{equation}
w_{\mbox{eff}}=-\frac{1}{3}\left(2x_3-1\right),~~q=1-x_3,~~\Omega_{\mbox {m}}=\frac{\kappa^2\rho_{\mbox m}}{3F'H^2}=1-x_1-x_2-x_3.
\label{30}
\end{equation}
For the particular case $\sigma =0$, we obtain from Eqs. (25) the function $m(r)=2r(r+1)$ ($r=x_3/x_2$). Replacing $m(r)$ into Eqs. (27) we obtain the functions of autonomous equations (26) \footnote{Our variable $x_2$ differs from that of Refs. Garcia-Salcedo et al. (2010) and Quiros and Urena-Lopez (2010).}
\[
f_1(x_1,x_2,x_3)=-1-x_3-3x_2+x_1^2-x_1x_3,
\]
\begin{equation}
f_2(x_1,x_2,x_3)=\frac{x_1x_2^2}{2(x_2+x_3)}-x_2\left(2x_3-4-x_1\right),
\label{31}
\end{equation}
\[
f_3(x_1,x_2,x_3)=-\frac{x_1x_2^2}{2(x_2+x_3)}-2x_3\left(x_3-2\right).
\]
The equilibrium points $\bar{x}_1$, $\bar{x}_2$, $\bar{x}_3$ are the solutions of the system of equations $f_i(x_1,x_2,x_3)=0$ ($i=1,2,3$). Replacing $x_i =\bar{x}_i + x'_i$ in Eq. (26), where $x'_i$ are the linear perturbations around the equilibrium points, one finds the evolution of $x'_i$ up to ${\cal O}(x'^{2}_i)$: $x'_i=c_{ij}\exp(\lambda_j N)$ (we imply a summation on repeated indexes), where $c_{ij}$ are constants and $\lambda_j$ are eigenvalues of
the Jacobian matrix $J_{ij}=\partial f_i(x_1,x_2,x_3)/\partial x_j$. From Eqs. (31), we obtain
equilibrium points, eigenvalues of the Jacobian matrix, and parameters defined in Eqs. (30) that are summarized in Table 2.
\begin{table}[ht]
\caption{Critical points ($\sigma =0$)}
\centering
\begin{tabular}{c c c c c c c c c c c c}
\hline \hline 
$P_i$ & $\bar{x}_1$ & $\bar{x}_2$ & $\bar{x}_3$ & $\Omega_{\mbox{m}}$ & $w_{\mbox{eff}}$ & $q$ & $r$ & $m$ & $\lambda_1$ & $\lambda_2$  &  $\lambda_3$ \\[0.5ex] 
\hline 
$P_1$ & 0 & -1 & 2 & 0 & -1 & -1 & -2 & 4 & -3 & $\frac{-3-\sqrt{21}}{2}$ &  $\frac{-3+\sqrt{21}}{2}$ \\
$P_2$ & -1 & 0 & 0 & 2 & 1/3 & 1 & -1 & 0 & -2 & $\frac{13-\sqrt{7}}{4}$ &  $\frac{13+\sqrt{7}}{4}$\\
$P_3$ & 1 & 0 & 0 & 0 & 1/3 & 1 & -1 & 0 & 2 & $\frac{19-\sqrt{7}}{4}$ &  $\frac{19+\sqrt{7}}{4}$\\
$P_4$ & -1 & 0 & 2 & 0 & -1 & -1 & $\infty$ & $\infty$ & -1 & -4 & -4\\
$P_5$ & 0 & -1/2 & 1/2 & 1 & 0 & 1/2 & -1 & 0 & div & div & div\\
$P_6$ & 3 & 0 & 2 & -4 & -1 & -1 & $\infty$ & $\infty$ & 3 & 4 & -4\\
$P_7$ & -3 & 2 & -1 & 3 & 1 & 2 & -1/2 & -1/2 & 3 & $\frac{-3-i\sqrt{15}}{2}$ & $\frac{-3+i\sqrt{15}}{2}$\\
[1ex] 
\hline 
\end{tabular}
\label{table:crit}
\end{table}

The equilibrium point $P_1$ corresponds to de-Sitter solutions because $w_{\mbox{eff}}=-1$, with accelerated
expansion ($q=-1$), and is a saddle point since one of the eigenvalues is positive. This point is unstable as $m(r=-2)>1$ (Amendola et al. (2007)).

The point $P_2$ corresponds to so-called $\phi$-matter dominated epoch ($\phi$MDE) (Amendola (2000)) and represents the ``wrong" matter epoch, $\Omega_{\mbox{m}}=2$. Eigenvalues $\lambda_2$, $\lambda_3$ are positive, and the point $P_2$ is a saddle equilibrium point with the decelerated expansion ($q=1$). It should be noted that calculating $\lambda_2$, $\lambda_3$ for the point $P_2$ (as well as for the point $P_3$), we use L'H\^{o}pital rule and imply that $x_2=x_3\rightarrow 0$. Since $w_{\mbox{eff}}=1/3$, this point mimics the radiation.

The fixed point $P_3$ is an unstable node because all eigenvalues $\lambda_i$ are positive. It corresponds to the past attractor for the decelerated epoch ($q=1$) and mimics the radiation ($w_{\mbox{eff}}=1/3$).

The point $P_4$ is similar to the point $P_1$ corresponding to de-Sitter solutions but it is a stable point ($\lambda_i<0$). This point mimics a cosmological constant ($w_{\mbox{eff}}=-1$) and can be the late-time
attractor.

The equilibrium point $P_5$ corresponds to a standard matter era ($a\propto t^{2/3}$) and the necessary condition (Amendola et al. (2007)) $m(r=-1)=0$ is satisfied, and the eigenvalues diverge.

The saddle fixed point $P_6$ does not correspond to any cosmological scenario since $\Omega_{\mbox {m}}<0$ but from the definition (see Eq. (30)) it should be positive.

The point $P_7$ is a saddle fixed point and corresponds to decelerated expansion ($q=2$). Since EoS parameter $w_{\mbox{eff}}=1$, this point mimics a stiff fluid.

\section{Discussion and Conclusion}

In recent paper (Kruglov (2015)), we have presented the model of $F(R)$ gravity with the function $F(R)=(1/\beta)\arcsin(\beta R)$ containing only one dimensional parameter $\beta$. There are some similarities and differences between two models. First of all the current model proposed contains two parameters $\lambda$ and $\sigma$ that allow us to adjust some observable. Thus, there is a set of constant curvature solutions (Table 1 in subsection 2.1), but in the model (Kruglov (2015)) we have only one constant curvature solution. If one looks at the Taylor series expansion of the function $F(R)$, it results in EH term $R$
plus odd powers of $R$ in the model (Kruglov (2015)), and the function (1) selected in this paper implies the EH term $R$ plus even powers. Hence, there are differences in the high curvature regimes in both models. Nevertheless, it is interesting that besides differences between $F(R)$ functions in two models, the parameters provided in both cases
are similar, which is evident from the figures presented in these papers (for a particular parameter $\sigma=-0.9$).
If one varies the parameter $\sigma$ in the model under consideration, we find differences in the behavior of the Universe evolution.

Thus, $F(R)$ gravity model proposed, with the Born$-$Infeld-like action, containing two scales ($\lambda$ and $\sigma$), describes inflation corresponding to the de Sitter solution. The bound on the constants $\lambda$, $\sigma$ from the local tests was evaluated, Eq. (4). The de Sitter spacetime is unstable and the constant curvature solution with zero Ricci scalar is stable. At small curvatures the action becomes the EH action and corresponds to GR without the cosmological constant. Thus, this model describes DE dynamically by the new gravitational physics. The Einstein frame was studied, the potential and the mass of the scalar degree of freedom were obtained (see Figs. 3 and 4). We have calculated the slow-roll parameters of the model, $\epsilon$, $\eta$, and ranges where they are small. It was demonstrated that corrections of $F(R)$ gravity model under consideration to GR are small. The critical points ($P_1$ and $P_5$ in the classification of the work of Amendola et al. (2007)) of autonomous equations are investigated and the function m(r) characterizing the deviation from the $\Lambda$CDM-model is calculated. Although the matter dominated epoch (the point $P_5$) in the model exists but the necessary conditions for the standard matter era are not satisfied ($m'(r=-1)=-2<-1$). As a result the particular model under examination can be an approximation describing the Universe evolution as many $F(R)$ gravity models considered in the literature. Even if this F(R) theory does not explain current acceleration of the Universe, it can describe early-time inflation and be alternative to GR.

\vspace{5mm}
\textbf{References}
\vspace{5mm}

Amendola, L., Gannouji, R., Polarski, D., Tsujikawa, S.: Phys. Rev. D \textbf{75}, 083504 (2007)
    (arXiv:gr-qc/0612180).

Amendola, L.: Phys. Rev. D \textbf{62}, 043511 (2000)
 (arXiv:astro-ph/9908023).

Appleby, S. A., Battye, R. A.: Phys. Lett. B \textbf{654}, 7 (2007)
 (arXiv:0705.3199).

Appleby, S. A., Battye, R. A., Starobinsky, A. A.: JCAP \textbf{1006}, 005 (2010)
(arXiv:0909.1737 [astro-ph.CO]).

Ba\~{n}ados, M., Ferreira, P. G.: Phys. Rev. Lett. \textbf{105}, 011101 (2010)
(arXiv:1006.1769 [astro-ph.CO]).

Barrow,  J. D., Ottewill, A. C.: J. Phys. A \textbf{16}, 2757 (1983)

Berry, C. P. L., Gair, J. R.: Phys. Rev. D \textbf{83}, 104022 (2011)
(arXiv:1104.0819 [gr-qc]).

Born,  M., Infeld, L.: Proc. R. Soc. A \textbf{144}, 425 (1934a).

Born,  M., Infeld, L.: Proc. R. Soc. A \textbf{147}, 522 (1934b).

Born,  M., Infeld, L.: A \textbf{150}, 141 (1935).

Comelli, D., Dolgov, A.: JHEP \textbf{0411}, 062 (2004)
(arXiv:gr-qc/0404065).

Comelli, D.: Phys. Rev. D \textbf{72}, 064018 (2005)
 (arXiv:gr-qc/0505088).

Deser, S., Gibbons, G. W.: Class. Quant. Grav. \textbf{15}, L35 (1998)
 (arXiv:hep-th/9803049).

Eingorn, M., Zhuk, A.: Phys. Rev. D \textbf{84}, 024023 (2011)
(arXiv:1104.1456 [gr-qc]).

Fabris, J. C. et al.: Phys. Rev. D \textbf{86}, 103525 (2012)
 (arXiv:1205.3458 [gr-qc]).

Fradkin, E. S., Tseytlin, A. A.: Phys. Lett. B \textbf{163}, 123 (1985).

Garcia-Salcedo, R. et al.: JCAP \textbf{1002}, 027 (2010) (arXiv:0912.5048 [gr-qc]).

Gates Jr., S. J., Ketov, S. V.: Class. Quant. Grav. \textbf{18}, 3561 (2001)
(arXiv: hep-th/0104223).

Gullu, I., Sisman, T. C., Tekin, B.: Phys. Rev. D \textbf{82}, 124023 (2010)
(arXiv:1010.2411 [hep-th]).

Harko, T., Lobo, F. S. N., Nojiri, S., and Odintsov, S. D.: Phys. Rev. D \textbf{84}, 024020 (2011)
 (arXiv:1104.2669 [gr-qc]). 

Herdeiro, C., Hirano, S., Sato, Y.: Phys. Rev. D \textbf{84}, 124048 (2011)
(arXiv:1110.0832 [gr-qc]).

Hu, W., Sawicki, I.: Phys. Rev. D \textbf{76}, 064004 (2007)
 (arXiv:0705.1158 [astro-ph]).

Kapner,  D. J. et al.: Phys. Rev. Lett. \textbf{98}, 021101 (2007)
(arXiv:hep-ph/0611184).

Kruglov, S. I.: J. Phys. A \textbf{43}, 375402 (2010)
 (arXiv:0909.1032 [hep-th]).

Kruglov, S. I.: Int. J. Theor. Phys. \textbf{52}, 2477 (2013)
 (arXiv:1202.4807 [gr-qc]).

Kruglov, S. I.: Astrophys. Space Sci. \textbf{358}, 48 (2015)
(arXiv: 1502.00659).

Liddle, A. R., Lyth, D. H.: Cosmological Inflation and Large-scale Structure,
Cambridge University Press, Cambridge (2000).

Magnano, G., Sokolowski, L. M.: Phys. Rev. D \textbf{50}, 5039 (1994)
 (arXiv:gr-qc/9312008).

Makarenko, A. N., Odintsov, S. D., Olmo, G. J., Rubiera-Garcia, D.: TSPU Bulletin \textbf{12}, 158 (2014) 
(arXiv:1411.6193 [gr-qc]).

M\"{u}ller,  V., Schmidt, H.-J., Starobinsky, A. A.: Phys. Lett. B \textbf{202}, 198 (1988)

N\"{a}f, J., Jetzer, P.: Phys. Rev. D \textbf{81}, 104003 (2010)
(arXiv:1004.2014 [gr-qc]).

Nieto, J. A.: Phys. Rev. D \textbf{70}, 044042 (2004)
(arXiv:hep-th/0402071).

Noureen, I. and Zubair, M.: Eur. Phys. J. C \textbf{75}, 62 (2015) (arXiv:1501.04484 [gr-qc]). 

Pani, P., Cardoso,  V., Delsate, T.: Phys. Rev. Lett. \textbf{107}, 031101 (2011)
 (arXiv:1106.3569 [gr-qc]).

Pleba\'{n}ski, J.: Lectures on non-linear electrodynamics, Nordita, Copenhagen (1970).

Quiros,  I., Leyva, Y., Napoles, Y.: Phys. Rev. D \textbf{80}, 024022 (2009)
(arXiv:0906.1190 [gr-qc]).

Quiros, I., Urena-Lopez, L. A.: Phys. Rev. D \textbf{82}, 044002 (2010)
(arXiv:1004.1719 [gr-qc]).

Sharif, M. and  Zubair, M.: Gen. Rel. Grav. \textbf{46}, 1723 (2014).

Starobinsky, A. A.: JETP Lett. \textbf{86}, 157 (2007)
 (arXiv:0706.2041).

Starobinsky, A. A.: Phys. Lett. B \textbf{91}, 99 (1980).

Wohlfarth, M. N. R.: Class. Quant. Grav. \textbf{21}, 1927 (2004a).

Wohlfarth, M. N. R.: Class. Quant. Grav. \textbf{21}, 5297 (2004b).
(arXiv:hep-th/0310067).

Zubair, M. and Abbas, G.: Astrophys. Space Sci. \textbf{357}, 154 (2015).

\end{document}